\documentclass[showpacs,oneside,twocolumn,prl,amsmath,amssymb,fleqn,superscriptaddress]{revtex4-1}
\usepackage{cases}
\usepackage{amsmath}
\usepackage{amssymb}
\usepackage{amsfonts}
\usepackage{amssymb}
\usepackage{dcolumn}
\usepackage{bm}
\usepackage{bbm}
\usepackage{graphicx}
\usepackage{xcolor}
\usepackage{array}
\usepackage{subfigure}
\usepackage[none]{hyphenat}

\newcommand{\bra}[1]{\langle #1\vert}
\newcommand{\ket}[1]{\vert #1\rangle}

\begin{document}
\title{An experimental test of the non-classicality of quantum mechanics using an unmovable and indivisible system}
\author{Xi Kong}
\affiliation{Hefei National Laboratory for Physics Sciences at
Microscale and Department of Modern Physics, University of Science
and Technology of China, Hefei, 230026, China}
\author{Mingjun Shi}
\affiliation{Hefei National Laboratory for Physics Sciences at
Microscale and Department of Modern Physics, University of Science
and Technology of China, Hefei, 230026, China}
\author{Fazhan Shi}
\affiliation{Hefei National Laboratory for Physics Sciences at
Microscale and Department of Modern Physics, University of Science
and Technology of China, Hefei, 230026, China}
\author{Pengfei Wang}
\affiliation{Hefei National Laboratory for Physics Sciences at
Microscale and Department of Modern Physics, University of Science
and Technology of China, Hefei, 230026, China}
\author{Pu Huang}
\affiliation{Hefei National Laboratory for Physics Sciences at
Microscale and Department of Modern Physics, University of Science
and Technology of China, Hefei, 230026, China}
\author{Qi Zhang}
\affiliation{Hefei National Laboratory for Physics Sciences at
Microscale and Department of Modern Physics, University of Science
and Technology of China, Hefei, 230026, China}
\author{Chenyong Ju}
\affiliation{Hefei National Laboratory for Physics Sciences at
Microscale and Department of Modern Physics, University of Science
and Technology of China, Hefei, 230026, China}
\author{Changkui Duan}
\affiliation{Hefei National Laboratory for Physics Sciences at
Microscale and Department of Modern Physics, University of Science
and Technology of China, Hefei, 230026, China}
\author{Sixia Yu}
\affiliation{Hefei National Laboratory for Physics Sciences at
Microscale and Department of Modern Physics, University of Science
and Technology of China, Hefei, 230026, China}
\affiliation{Centre for quantum technologies, National University of Singapore, 3
Science Drive 2, Singapore 117543, Singapore}
\author{Jiangfeng Du}
\altaffiliation{djf@ustc.edu.cn}
\affiliation{Hefei National Laboratory for Physics Sciences at
Microscale and Department of Modern Physics, University of Science
and Technology of China, Hefei, 230026, China}

\begin{abstract}

Quantum mechanics provides a statistical description about nature, and thus would be incomplete if its statistical predictions could not
be accounted for by some realistic models with hidden variables. There are, however, two powerful theorems against the hidden-variable theories showing that certain quantum features cannot be reproduced based on two rationale premises of locality, Bell's theorem, and noncontextuality, due to Bell, Kochen and Specker (BKS). Noncontextuality is independent of nonlocality, and the contextuality manifests itself even in a single object. Here we report an experimental verification of quantum contextuality by a single spin-1 electron system at room temperature. Such a three-level system is indivisible and then we close the compatibility loophole which exists in the experiments performed on bipartite systems. Our results confirm the quantum contextuality to be the intrinsic property of single particles.

\end{abstract}


\maketitle

In quantum mechanics, not all properties can be simultaneously well defined.
Such incompatibility of properties, characterized by Heisenberg uncertainty principle,
is one of the most curious and surprising features of quantum mechanics,
and conflicts strongly with our experience in daily lives.
Hidden variable (HV) theory aims at extending quantum mechanics into a more fundamental theory
which provides a classical-like deterministic description of the nature.
An intuitive feature of the classical description is { its non-contextuality}:
the result of a measurement of an observable is predetermined and independent of which set of compatible (i.e., commeasurable) observables might be measured alongside.
Namely, if $A$, $B$ and $C$ are observables such that $A$ and $B$ commute, $A$ and $C$ commute,
but $B$ and $C$ do not commute, then the value predicted to occur in a measurement of $A$
does not depend on whether $B$ or $C$ was measured simultaneously.
The theorems derived by Bell \cite{No2}, Kochen and Specker \cite{No3}, called BKS theorems, shows that non-contextuality hidden variable (NCHV) is in conflict with quantum mechanics.

To confirm the theorems, many theoretical schemes \cite{No4,No5,No6,No7,No8,No9,No10} have been proposed for possible experimental tests of quantum contextuality. Unfortunately, it had been a conundrum for experimentalists to accomplish such a test, because the BKS theorems were ¡°nullified¡± in real experiments due to the unreachable measurement precision \cite{No11}. Not until recently it was shown that the BKS theorems could be converted into experimentally available schemes by correlations between compatible measurements based on some inequalities called non-contextuality inequalities \cite{No12,No13,Yu2012prl}.

The previous experimental tests of the BKS theorems were performed using \cite{No14,No15,No16} and neutrons \cite{No17,No18}.
Recently, experiments using ions \cite{No19} and NMR \cite{No20} have been accomplished.
Although the results obtained in above-mentioned experiments were completely in conflict with non-contextuality,
the involvement of at least two particles in those experiments left some loopholes open, such as the uncontrollable inter-particle interactions probably reducing the compatibility of the measured observables, and the detection loophole in multi-photon experiments due to photon loss and phase instability.
In this sense, the experiments performed on single-particle systems are more compelling and highly desired, but also of great challenge with currently available technology.
Moreover, since tens of observables are required to accomplish the proof of BKS theorem for a spin-1 particle \cite{No10,No21}, it is an insurmountable obstacle for experimentalists to find a qualified system to measure the observables precisely.
The latest experiments demonstrating the conflict with NCHV theories were implemented using single photons \cite{No22,ZuarXiv1207.0059} and ions \cite{ZhangarXiv1209.3831v1}, rather than unmovable and indivisible solid-state systems.

 In this letter, we report an experimental realization of a true single-particle verification of the quantum contextuality by measuring five properly chosen observables (given below, see \cite{No23} for more details) in a spin-1 system. Since it is a single-particle-oriented experiment, we prevent a compatibility loophole \cite{Guhne2010pra} regarding inter-particle interactions or entanglement. Moreover, we only use the individual, rather than correlated, measurement in our implementation. Neither photon interference nor coincidence record is needed. We show below that our experiment enables a precise detection of a small violation of the non-contextuality inequality.

To begin with, we briefly outline the theoretical scheme that excludes NCHV models for a spin-1 system.
In any NCHV model we consider five observables $L_i$ taking values in $\{0,1\}$ with $i=1,\cdots, 5$. Suppose that observables $L_i$ and $L_{i+1}$ are compatible, { with identification $L_6=L_1$, then it holds
\begin{equation}\label{NCHV inequality}
\left ( \sum_{i=1}^{5}\langle L_i\rangle-\sum_{i=1}^{5} \langle L_i L_{i+1}\rangle\right )_{\mathrm{NCHV}} \leqslant 2.
\end{equation}
Here the expectation values are taken with respect to certain probability distribution of hidden variables that determine the values of $L_i$'s. Alternatively, Cabello and co-workers \cite{CabelloArXiv1010.2163} argued that the second term in the above inequality can be dropped, { conditioned on the compatibility or the orthogonality of projections representing observables $L_i$ and $L_{i+1}$, i.e.,}
\begin{equation}\label{NCHV_Cabello}
\sum_{i=1}^{5}\langle L_i\rangle_{\mathrm{NCHV}} \leqslant 2.
\end{equation}

In the quantum mechanical description of a spin-1 system we denote by $S_{\ell_j}^2$ the square of the spin operator along the unit vector $\ell_j$.
If two unit vectors $\ell_i$ and $\ell_j$ are orthogonal, then the observables $S_{\ell_i}^2$ and $S_{\ell_j}^2$ are compatible and can be measured simultaneously.
We consider a cyclic quintuplet of unit vector with $\ell_i\perp\ell_{i+1}$
with $i=1,\cdots, 5$ and $\ell_6=\ell_1$ and five
corresponding observables $L_i=\mathbbm{1}-S_{\ell_i}^2=\ket{\ell_i}\bra{\ell_i}$ with $\mathbbm{1}$ the $3\times3$ identity matrix and they are cyclic compatible.
Here $\ket{\ell_i}$, called neutrally polarized state,
is the eigenstate of $S_{\ell_i}$ with eigenvalues $0$.
The form of $\ket{\ell_i}$ is just the same as the unit vector $\ell_i$ in three-dimensional real space.
The expectations $\langle L_i\rangle$ are calculated or measured with respect to certain state $\ket{\psi}$ of the particle.
The state $\ket{\psi}$ is chosen as the neutrally polarized state directed along the fivefold symmetry axis of the regular pentagram, the vertices of which correspond to the five vectors $\ell_i$.
Then $\langle L_i\rangle_{\psi}=\bra{\psi}L_i\ket{\psi}=|\bra{\psi}\ell_i\rangle|^2=1/\sqrt{5}$ and as a result of cyclical orthogonality, $\langle L_i L_{i+1}\rangle_{\psi} =  \bra{\psi}\ell_i\rangle \bra{\ell_i}\ell_{i+1}\rangle \bra{\ell_{i+1}} \psi\rangle = 0$.
It follows that
\begin{eqnarray}
&& \sum_{i=1}^{5}\langle L_i\rangle_{\psi} - \sum_{i=1}^{5} \langle L_i L_{i+1}\rangle_{\psi} \nonumber \\
&=& \sum_{i=1}^{5}\langle L_i\rangle_{\psi}  =\sqrt{5}{>} 2.
\end{eqnarray}
This means that quantum mechanical prediction violates the NCHV inequality \eqref{NCHV inequality}.
We show below the experimental observation of this violation.}

\begin{figure}[htbp]
\centering
\includegraphics[width=1\columnwidth]{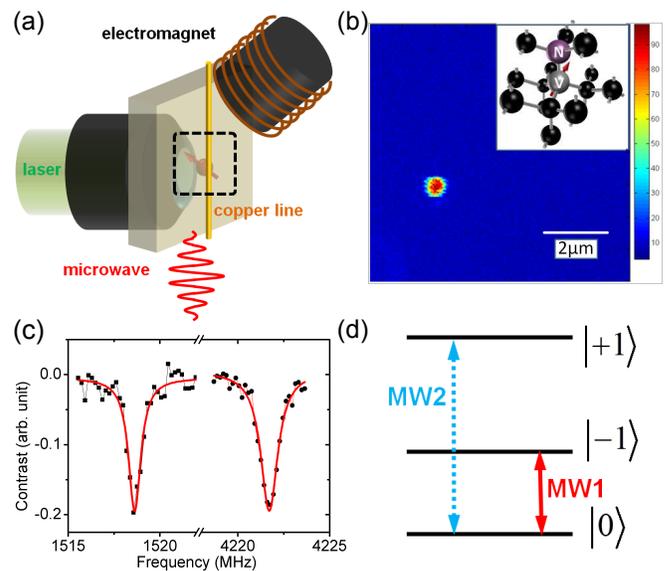}

\caption{(Color online). Single particle three-level qutrit system. (a) Schematic the confocal microscopy setup. The state of qutrit was initialized and readout by green laser illumination through the objective. An external DC magnetic field was applied by a movable electric magnet while microwave was carried by coplanar wire with 20 $\mu$m diameter. (b) Fluorescence of confocal microscopy image of a single NV defect. The inset presents the structure of the NV defect in diamond. (c) The electron spin resonance spectrum of the NV centre, which shows energy lever splitting of the electron spin. The left peak is 1518.6 MHz, corresponding to the transition between $\ket{0}$ and $\ket{-1}$. The right peak is 4221.7 MHz, corresponding to the transition between $\ket{0}$ and $\ket{1}$. (d) The electron energy level scheme of the NV defect for the case of nitrogen nuclear spin being polarized, where the red solid and blue dash lines denote the transitions for the lower frequency microwave pulse (MW1) and the higher frequency microwave pulse (MW2), respectively.
}
 \label{fig1}
\end{figure}

In our experiment, we employed the spin-1 qutrit based on a single negatively charged nitrogen-vacancy (NV) center in diamond. The NV center was consisted by an impurity nitrogen and a neighbour vacancy [the insert of FIG.~\ref{fig1}(b)]
which located in type-IIa bulk diamond with nitrogen concentration less than $5$ ppb and 1.1~\% natural abundance $^{13}$C.
The well-known Hamiltonian of the NV center in a static magnetic field $\bm{B}$ is given by
$H=DS_z^2+g_e\beta_e\bm{S}\cdot \bm{B}$, where $g_e$ is the electronic $g$-factor and $\beta_e$ is the Bohr magneton.
The vector $\bm{S}$ is the operator for the electron spin.
$D$, the zero field splitting, is equal to $2870$ MHz. A $482.7$~Gauss magnetic field was applied along the crystal axis $ \langle 111 \rangle$, and the energy splitting linearly depended on the external magnetic field magnitude. Individual NV centers were optically addressed by a confocal microscope mounted on a piezoelectric nano-scanner. FIG.~1(b) shows the scan map and the structure of NV center. Optically detected magnetic resonance spectrum was measured [FIG.~\ref{fig1}(c)] to determine the energies for the electron spin-1 levels shown in FIG.~\ref{fig1}(d), where eigenstates $\ket{m_e=0}$ and $\ket{m_e=\pm1}$ are denoted in short by $\ket{0}$ and $\ket{\pm1}$. { The required} two channels resonance microwave pulses were applied to the NV center for state manipulation.

\begin{figure}[htbp]
\centering
\includegraphics[width=1\columnwidth]{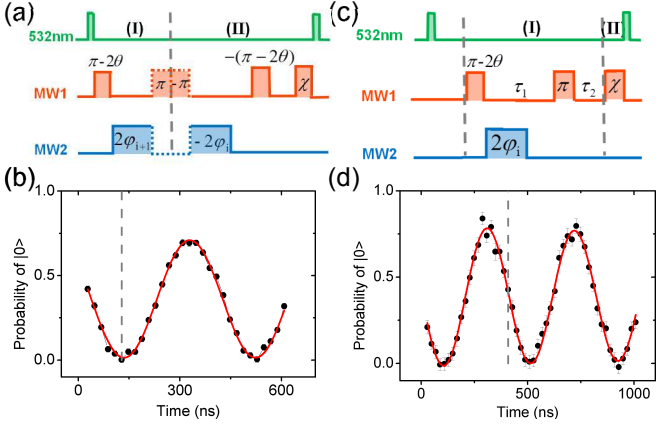}
\caption{(Color online). {Pulse sequences for state-overlap test and noncontextuality measurements.} (a) the pulse sequence to check the orthogonality between $|\ell_i\rangle$ and $|\ell_{i+1}\rangle$ ($i=1, \cdots,~5$). The minus sign `$-$' before the pulse duration time means a {180$^{\circ}$} phase shift from normal pulses. (b) The square module of the overlap obtained, where the intersection of the vertical dash line and the sine curve represents the orthogonality. For simplicity in experiment, we ignore the unit operation marked with the dashed pulses and short the MW2 pulse from {`2$\varphi _{i+1}$' and `$-2\varphi _i$'} to 2($\varphi _{i+1}-\varphi _i$). (c) The pulse sequence to measure $|\langle \psi |\ell_i\rangle|^2$ ($i=1, \cdots,~5$) and the result for $|\ell_5 \rangle$ was shown in (d). The data were collected on the time $\chi = 2\pi$ instead of $\chi = 0$ for technical issue. $\tau _1$ = $\tau _2$ was set for depressing the effect of the noise in the solid system.
}
 \label{fig2}
\end{figure}

The five cyclically orthogonal states $\ket{\ell_n}$ for $\ell_n = (\sin\theta\cos\varphi_n,\;\sin\theta\sin\varphi_n,\;\cos\theta)$,
where $\varphi_n = 0.8 (n-1) \pi~(\mod 2\pi)$ ($n=1-5$), and $\theta=\arccos(5^{-1/4})$, are prepared using the pulse sequence given in the stage (1) of FIG.~\ref{fig2}(a) for $n=i+1$  (and also FIG.~\ref{fig2}(c) for $n=i$). Due to the compatibility requirement of Bell-KS theorems, the observables need to be cyclically commutable. Otherwise the hidden variables are disturbed during observation and the ¡®compatibility loophole¡¯ \cite{No24} is potentially opened. For spin-1 system, $[S_{l_i}^2,S_{l_{i+1}}^2] = 0$ is equivalent to $\ell_i \perp \ell_{i+1}$. Hence we need only to perform a test to check the orthogonality between two successive $\ket{\ell_n}$'s. This is done by using the pulse sequences given in FIG.~\ref{fig2}(a), where stage (I) and (II) realized the evolution of $\ket{\ell_{i+1}}$ from $\ket{0}$ and the inverted evolution of $\ket{\ell_i}$ to $\ket{0}$.  Then optical readout enabled the observation of $|\langle \ell_i|l_{i+1}\rangle |^2$.
As shown in FIG.~\ref{fig2}(b), the measured average overlap $\overline{|\langle \ell_i|\ell_{i+1}\rangle |^2} = 0.0020\pm 0.0061$, confirmed the compatibility of the observables. Hence, the compatibility loophole is decisively closed in our experiment.

FIG.~\ref{fig2}(c) shows the pulse sequences used to test the quantum contextuality. The state $\ket{\ell_i}$ prepared in the stage (I) of FIG.~\ref{fig2}(c) is detected by measuring the fluorescence $I_{\rm PL}$ after applying a $\chi$ pulse for $\ket{0} \Leftrightarrow \ket{-1}$. The probability of the state $\ket{\ell_i}$ being measured as $\ket{0}$ is then derived and the result is plotted in FIG.~\ref{fig2}(d) as a function of the pulse length of $\chi$. The point where $\chi = 0$ (mod $2\pi$) in FIG.~\ref{fig2}(d) corresponds to the case of $\langle L_i\rangle_{\psi} = \langle \psi |L_i |\psi \rangle = |\langle \psi|\ell_i\rangle|^2$, where $\ket{\psi} = \ket{0}$.

\begin{figure}[htbp]
\centering
\includegraphics[width=1\columnwidth]{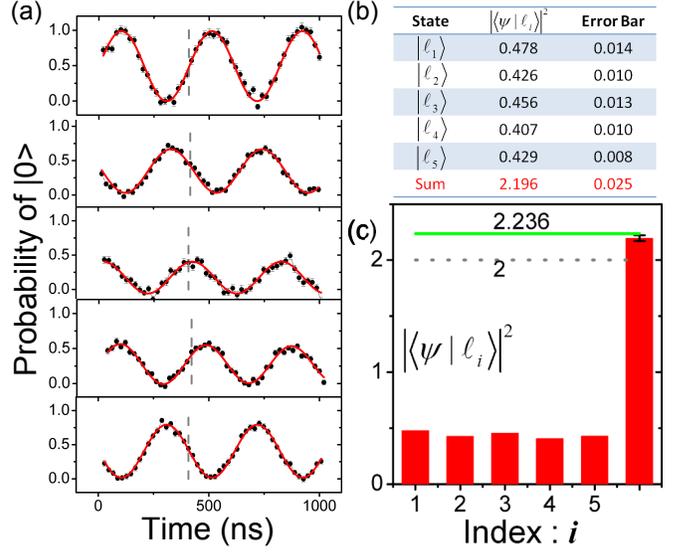}
%\vskip 3cm
\caption{(Color online). Experimental results for $|\langle \psi |\ell_i\rangle|^2$ ($i=1,~\cdots,~5$).
(a) The state-overlap curve results to determine $|\langle \psi |\ell_i\rangle|^2$ ($i=1,~\cdots,~5$) from upper curve to bottom. The vertical dashed lines show the place where the data are designated to be read off. (b) The measurement results are listed. The data were collected from the fitting curve and the error bars were calculated with $\sqrt {{{(\delta A)}^2} + {{(\delta {y_0})}^2}} $, in which $\delta A$ and $\delta y_0$ are the fitting error of amplitude and offset on the measurement curves. (c) The measurement result shown in histogram. Compared to theoretical predictions, the sum exceeds the upper limit allowed by HV models. The solid green and dashed gray lines show the prediction of quantum mechanics and the upper limit of NCHV theories.
\label{fig3}
}
\end{figure}

The measurement of $|\langle \psi |\ell_i\rangle |^2$  with $\ket{\psi} =\ket{0}$ for all the five states are shown in FIG.~3. By these measurements and the cyclical orthogonality of $\ket{\ell_i}$, we obtain the following results.
{
\begin{eqnarray}
\sum\limits_{i=1}^5 \langle L_i\rangle _{\psi} = 2.196~(\pm 0.025) > 2.
\end{eqnarray}}%
This result demonstrates the violation of the NCHV inequality { \eqref{NCHV_Cabello}} by about eight standard deviations and agrees well with the quantum mechanical expectation.

In the above experiments we have avoided the troublesome correlation measurements by resorting to, as a result of compatibility between $L_i$ and $L_{i+1}$ ($i=1,~\cdots,~5$), the orthogonality $\bra{\ell_i}\ell_{i+1}\rangle=0$. In a realistic measurement $\bra{\ell_i}\ell_{i+1}\rangle=0$ may not strictly hold.
In the worst scenario all the directions $\ell_i$ are along the direction of the state $|\psi\rangle$ then one can easily get an expectation value as large as 5. In fact in our experiment we have the average { $\epsilon=\overline {\bra{\ell_i}\ell_{i+1}\rangle|^2} = \sum_i|\bra{\ell_i}\ell_{i+1}\rangle|^2/5=0.0020$}. Therefore a quantitative trade-off between the orthogonality and the effective violation, i.e., the violation lead to the conclusion that quantum mechanics is contextual, should be in order.

Consider the correlation of two observable $L_i=|\ell_i\rangle\langle \ell_i|$ and $L_{i+1}=|\ell_{i+1}\rangle\langle \ell_{i+1}|$ in an arbitrary state $\varrho$ we have $|\langle L_iL_{i+1}\rangle_{\varrho}|=|\langle\ell_i|\ell_{i+1}\rangle||\langle \ell_i|\varrho|\ell_{i+1}\rangle|$. Since $|\langle \ell_i|\varrho|\ell_{i+1}\rangle|=|\langle \ell_i|\sqrt\varrho\sqrt\varrho|\ell_{i+1}\rangle|\le \sqrt{\langle L_i\rangle\langle L_{i+1}\rangle}$, we have
{
\begin{eqnarray}
&&\sum_{i=1}^{5}\langle L_i\rangle_{\psi} - \sum_{i=1}^{5} |\langle L_i L_{i+1}\rangle_{\psi}| \nonumber \\ &\geqslant & \sum_{i=1}^{5}\langle L_i\rangle_{\psi} \nonumber
-\sum_{i=1}^5 |\langle\ell_i|\ell_{i+1}\rangle|\sqrt{\langle L_i\rangle_{\psi}\langle L_{i+1}\rangle_{\psi}}\\
&\geqslant&\sum_{i=1}^{5}\langle L_i\rangle_{\psi}
-\sqrt{5\epsilon\sum_{i=1}^5 \langle L_i\rangle_{\psi}\langle L_{i+1}\rangle_{\psi}}= 2.098
\end{eqnarray}}%
according to the experimental data. In other words, our experiment gives a lower bound of the quantum mechanical prediction of the left-hand side of non-contextuality
inequality Eq.~(\ref{NCHV inequality}) and still we have a violation.

There are two main factors possibly damaging the validity of our experimental results, i.e., the hyperfine interaction between the electron spin and the nearby nitrogen nuclear spin, and decoherence due to environment.
In our employed NV center system, the Hamiltonian between NV and the nearby nitrogen nuclear spin is $\bm{S}\cdot \mathbb{A}\cdot\bm{I}-g_n\beta_n\bm{I}\cdot\bm{B}$, where $g_n$ is the nuclear $g$-factor and $\beta_n$ is the nuclear magneton.
$\bm{S}$ and $\bm{I}$ are operators associated with electronic and nuclear spins, respectively.
$\mathbb{A}$ is a tensor relevant to the hyperfine interaction with splitting in $z$ axis about $2.2$ MHz.
A magnetic field of $482.7$ Gauss was used to polarized the nitrogen nuclear spin \cite{Wrachtrup 2009 PRL, Wrachtrup 2010 PRB}, so that we can work within the three-state subspace of a given $m_i$=1. Since we employed a sufficiently large Rabi frequency of MW2 to achieve manipulation within a much shorter time compared to the decoherence time, so the drift can be omitted.

As for the decoherence in our operations, the decay time $T^{\prime}_2$ is roughly 35 $\mu$s in the step of preparing the state $|\ell_i\rangle$ which is fitted from the decay of Rabi oscillation due to static magnetic-field inhomogeneities or other instrumental imperfections \cite{No28}.
In the step of generating the remaining states, the error mainly comes from the decoherence due to the ${}^{13}$C spin bath \cite{No25}. The coherence time is $T_2 = (148 \pm 17)~\mu$s. Our experiment was limited by T$_2$ for the Hahn echo sequence was employed in our measurement sequences. Due to these imperfection factors, $|\langle \psi|\ell_i\rangle |^2$ is slightly deviated from its ideal value due to the small drift less than $0.02$ in $I_{\rm PL}$ for $P = 0$.
This also implies that a more evident violation of the inequality could be obtained under perfect manipulation.

In summary, our observation has demonstrated unambiguously the violation of the non-contextuality inequality, which cannot be explained by any non-contextual model. Remarkably, our implementation, performed on a single particle in solid state, has closed the compatible loophole which exists in the experiments performed on bipartite systems, and is therefore more precise and convincing.  Further improvement can be done to implement detection-loophole-free tests. Our experiment may give profound impacts on quantum mechanical foundation, and we expect this result to stimulate more elegant experiments for further exploration of the peculiar characteristic of quantum physics.

This work was supported by the National Key Basic Research Program of China (Grant No. 2013CB921800), the National Natural Science Foundation of China (Grants Nos. 10834005, 91021005, 11275183, 11104262 and 11161160553), the Instrument Developing Project of the Chinese Academy of Sciences (Grant No. Y2010025), and the National Research Foundation and Ministry of Education, Singapore (Grant No. WBS: R-710-000-008-271).

\end{document}